\documentclass{article}

\usepackage[utf8]{inputenc} 
\usepackage[T1]{fontenc}    
\usepackage{hyperref}       
\usepackage{url}            
\usepackage{booktabs}       
\usepackage{amsfonts}       
\usepackage{nicefrac}       
\usepackage{microtype}      
\usepackage{geometry}

\usepackage{subfig}
\usepackage{bm}
\usepackage{tikz}
\usetikzlibrary{matrix}
\usepackage{array}
\usepackage{multirow}
\usepackage{verbatim}
\usepackage{amsmath}
\usepackage{amssymb}
\usepackage{amsfonts}
\usepackage{navigator}

\newcommand{\walderurl}{http://bit.ly/1PqNTJ2}

\makeatletter
\newcommand{\thickhline}{%
    \noalign {\ifnum 0=`}\fi \hrule height 1pt
    \futurelet \reserved@a \@xhline
}
\newcolumntype{"}{@{\hskip\tabcolsep\vrule width 1pt\hskip\tabcolsep}}
\makeatother

\newcommand{\chapternewpage}{~ \ifodd\value{page} \chapter*{} \fi}

\newcommand{\novelty}[3]{\ifthenelse{\equal{\value{#1}}{0}}{#3}{#2}\setcounter{#1}{1}}

\newcommand{\beqn}{\begin{equation}\/}
\newcommand{\eeqn}{\end{equation}\/}
\newcommand{\beqns}{\begin{equation*}\/}
\newcommand{\eeqns}{\end{equation*}\/}
\newcommand{\beqna}{\begin{eqnarray}\/}
\newcommand{\eeqna}{\end{eqnarray}\/}
\newcommand{\beqnas}{\begin{eqnarray*}\/}
\newcommand{\eeqnas}{\end{eqnarray*}\/}

\newcommand{\balign}{\begin{align}\/}
\newcommand{\ealign}{\end{align}\/}
\newcommand{\bals}{\begin{align*}\/}
\newcommand{\eals}{\end{align*}\/}

\newcommand{\lnb}{\left(}
\newcommand{\rnb}{\right)}

\newcommand{\lcb}{\left\{}
\newcommand{\rcb}{\right\}}

\newcommand{\myvec}{\bm}

\newcommand{\cvec}{\myvec{c}}

\newcommand{\ovec}{\myvec{o}}

\newcommand{\svec}{\myvec{s}}

\newcommand{\xvec}{\myvec{x}}

\newcommand{\Ccal}{\mathcal{C}}

\newcommand{\Ocal}{\mathcal{O}}

\newcommand{\Xcal}{\mathcal{X}}

  {\begin{center}\begin{Sbox}\begin{minipage}}%
  {\end{minipage}\end{Sbox}\fbox{\TheSbox}\end{center}}

  {\begin{center}\begin{Sbox}\begin{minipage}}%
  {\end{minipage}\end{Sbox}\Ovalbox{\TheSbox}\end{center}}

\newcommand{\cmcolor}{}

\title{Modelling Symbolic Music: Beyond the Piano Roll}

\author{
  Christian Walder \\
  CSIRO Data61 \\
  7 London Circuit, Canberra, 2604, Australia. \\
  \texttt{christian.walder@data61.csiro.au} \\
}

\begin{document}
\maketitle

\begin{abstract}
In this paper, we consider the problem of probabilistically modelling symbolic music data. We introduce a representation which reduces polyphonic music to a univariate categorical sequence. In this way, we are able to apply state of the art natural language processing techniques, namely the long short-term memory sequence model. The representation we employ permits arbitrary rhythmic structure, which we assume to be given.  We show that our model is effective on four out of four piano roll based benchmark datasets. We further improve our model by augmenting our training data set with transpositions of the original pieces through all musical keys, thereby convincingly advancing the state of the art on these benchmark problems. We also fit models to music which is unconstrained in its rhythmic structure, discuss the properties of this model, and provide musical samples which are more sophisticated than previously possible with this class of recurrent neural network sequence models. We also provide our newly preprocessed data set of non piano-roll music data.
\end{abstract} 

\section{Introduction}
\label{sec:introduction}

Algorithmic music composition is an interesting and challenging problem which has inspired a wide variety of investigations \cite{aimusicsurvey}. A number of factors make this problem challenging. From a neuroscience perspective, there is a growing body of evidence which links the way we perceive music and the way we perceive natural language \cite{patelbook}. As such, we are highly sensitive to subtleties and irregularities in either of these sequential data streams. 

The intensity of natural language processing (NLP) research has surged along with the availability of datasets derived from the internet. Within the field of NLP, the broad range of techniques collectively known as deep learning have proven dominant in a number of sub-domains --- see \textit{e.g.} \cite{billionword}. Much of this research may find analogy in music, including algorithmic composition. Indeed, the very problem of data driven natural language generation is currently the subject of vigorous investigation by NLP researchers \cite{gravesgenerating, senticap}. 

Nonetheless, algorithmic composition presents unique technical challenges. In contrast to the words that make up text data, musical notes may occur contemporaneously. In this way, viewed as a sequence modelling problem, music data is multivariate. Moreover, the distribution of valid combinations of notes is highly multi-modal --- while a number of musical chords may be plausible in a given context, arbitrary combinations of the notes which make up those chords may be rather implausible.

The outline of the paper is as follows. In \autoref{sec:relatedwork} we begin by providing an overview of the relevant work in NLP and algorithmic composition. Section \ref{sec:approach} introduces the key elements of our approach: Long Short-Term Memory (LSTM) sequence modelling, our reduction to univariate prediction, our data representation, and our data augmentation scheme. In  \autoref{sec:experiments} we present our experimental results, before summing up and suggesting some possible directions for future work in the final \autoref{sec:summary}.

\section{Related Work}
\label{sec:relatedwork}

\subsection{Natural Language Processing}
\label{sec:nlp}

Early NLP work focused on intuitively plausible models such as Chomsky's (arguably Anglo-centric) context free grammars, and their probabilistic generalizations. More recently however, less intuitive schemes such as \textit{n-gram} models have proven more reliable for many real world problems \cite{manningbook}. In just the last few years, an arguably even more opaque and data driven set of techniques collectively known as deep learning have been declared the new state of the art on a range of tasks \cite{billionword}. A key building block is the so called recurrent neural network (RNN), the example of which we are most concerned with is the LSTM of Hochreiter and Schmidhuber \cite{lstm}. Like all RNNs, the LSTM naturally models sequential data by receiving as one of its inputs its own output from the previous time step. As clearly articulated in the original work \cite{lstm}, the LSTM is especially designed to overcome certain technical difficulties which would otherwise prevent the capturing of long range dependencies in sequential data. Nonetheless, the power of the model has been largely dormant until the ready availability of massively parallel computing architectures in the form of relatively inexpensive graphics processing units (GPUs). 
In a basic LSTM language model, the output $\ovec_t$ of the LSTM at time $t$ is used to predict the next word $\xvec_{t+1}$, while the unobserved state $\svec_t$ stores contextual information. The $\xvec_t$ typically encode words as \textit{one hot} vectors with dimension equal to the size of the lexicon and a single non-zero element equal to one at the index of the word. Before inputting to the LSTM cell, the $\xvec_t$ are transformed linearly by multiplication with an \textit{embedding matrix}, which captures word level semantics. This embedding may be learned in the context of the language model, or \textit{pre-trained} using a dedicated algorithm such as \cite{glove}. 

\subsection{Algorithmic Composition}
\label{sec:algomusic}

Polyphonic music modelling may be viewed as a sequence modelling problem, in which the elements of the sequence are, for example, those sets of notes which are sounding at a particular instant in time. Sequence modelling has a long history in machine learning, and a large body of literature exists which studies such classic approaches as the hidden Markov model (HMM) \cite{bishopbook}.  Recently, more general RNN models have been shown to model musical sequences more effectively in practice \cite{bl}. 

RNNs have long been applied to music composition. The CONCERT model of Mozer \cite{mozer1994} models a melody line along with its rhythmic structure, as well as an accompanying chord sequence. CONCERT uses the five dimensional pitch representation of Shephard \cite{shepard} in order to capture a notion of perceptual distance. Another interesting feature of CONCERT is the treatment of rhythmic information, in which note durations are explicitly modeled. Aside from CONCERT, the majority of models employ a simple uniform time discretization, which severely limits the applicability of the models to realistic settings.

Allan and Williams used an HMM with hidden state fixed to a given harmonic sequence \cite{moray}, and learned transition probabilities (between chords) and emission probabilities (governing note level realizations).

The suitability of the LSTM for algorithmic composition was noted in the original work \cite{lstm}, and later investigated by Eck and Schmidhuber for the blues genre \cite{eckschmidhuber}, although the data set and model parameter space were both small in the context of current GPU hardware.

Finally, we note that other ideas specifically from NLP have also been used for modelling music, including hierarchical grammars \cite{granrothgrammar,harmtrace}.

\section{Our Approach}
\label{sec:approach}

\subsection{LSTM Language Model Architecture}

We wish to leverage insights from state of the art language models, so we take a particular multi-layer LSTM as our starting point \cite{rnnregularization}. This model broadly fits our general description of the LSTM based language model in \autoref{sec:nlp}, with several non-trivial computational refinements such as an appropriate application of \textit{dropout} based regularization \cite{dropout}. We will show that the model of \cite{rnnregularization} is effective in our context as well, by reducing the music modelling problem to a suitable form.

\subsection{Reduction to Univariate Prediction}
\label{sec:reduction}


\begin{figure*}%
  \centering
  \subfloat[][staff notation]{
  \includegraphics[width=0.25\textwidth]{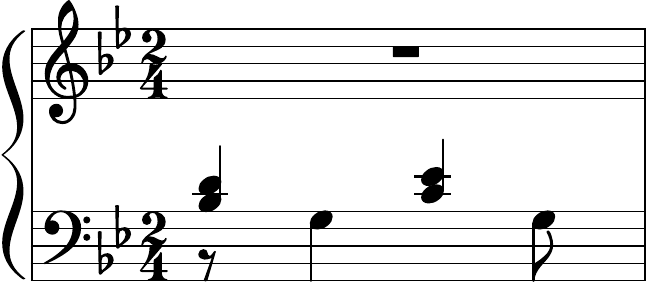}
  }
  \qquad
  \subfloat[][our representation]{
  {
  \resizebox{0.6\textwidth}{!}{%
\begin{tabular}{c||c|c|c|c|c|c|c|c|c|c}
midi 63 & 0 & 0 & 0 & 0 & 0 & 0 & 1 & 1 & 1 & 0 \\
midi 62 & 0 & 0 & 1 & 1 & 0 & 0 & 0 & 0 & 0 & 0 \\
midi 60 & 0 & 0 & 0 & 0 & 0 & 1 & 1 & 1 & 1 & 0 \\
midi 58 & 0 & 1 & 1 & 1 & 0 & 0 & 0 & 0 & 0 & 0 \\
midi 55 & 0 & 0 & 0 & 1 & 1 & 1 & 1 & 0 & 1 & 0 \\
$t$ & 0 & 0 & 0 & 0.5 & 1 & 1 & 1 & 1.5 & 1.5 & 2 \\
$\Delta t_{\text{event}}$ & 1 & 1 & 1 & 0 & 1 & 1 & 0 & 0.5 & 0 & 0 \\
$\Delta t_{\text{step}}$ & 0 & 0 & 0.5 & 0.5 & 0 & 0 & 0.5 & 0 & 0.5 & 0 \\
$\epsilon_{\text{onset}}$ & 0 & 1 & 0 & 0 & 0 & 1 & 0 & 0 & 0 & 0 \\
$\epsilon_{\text{offset}}$ & 0 & 0 & 0 & 0 & 1 & 0 & 0 & 1 & 0 & 1 \\
\hline
target & 70 & 74 & 67 & - & 72 & 75 & - & 67 & - & - \\
lower bound & 0 & 70 & 0 & - & 0 & 72 & - & 0 & - & - \\
\end{tabular}
}

  }
  }
  \caption{The first bar of Tchaikovsky's \textit{Song of the Lark} --- see section \ref{sec:representation} for a description.
  \label{fig:ty_maerz}
  }
\end{figure*}

It is possible, as in the RNN-Restricted Boltzmann Machine (RNN-RBM) of \cite{bl}, to modify the \textit{output} or \textit{prediction} layer of the basic  RNN language model architecture, such that it supports multivariate prediction. As we shall see, our reduction to a simple univariate problem is at least as effective in practice. Moreover, by reducing the problem in this way, we can easily leverage future advances in the highly active research area of NLP.

The reduction is motivated by the problem of modelling a density over multisets $\cvec\in\Ccal$ with underlying set of elements $\Xcal=\lcb 1, 2, \ldots, d \rcb$. For our purposes, $\Xcal$ represents the set of $d$ musical notes (or piano keys, say), and $\cvec$ is an unordered set of (possibly repeated) notes in a chord. Since $\Xcal$ contains ordinal elements, we can define a bijective function $f:\Ccal\rightarrow\Ocal$
, where $\Ocal$ is the set of ordered tuples with underlying set of elements $\Xcal$. Now, for $\ovec\in\Ocal$ such that $\ovec=\lnb o_1 \leq o_2 \leq \ldots, o_{|\ovec|} \rnb$, we can always write
\begin{equation}
\label{eqn:tuplefactorisation}
p(\ovec) = \prod_{i=1}^{|\ovec|} p(o_i | o_1, o_2, \ldots , o_{i-1}).
\end{equation}
Hence, we can learn (or infer) $p(\cvec)$ by solving a sequence of univariate learning problems. If the cardinality $|\ovec|$ is unknown, we simply model an ``end'' symbol, which we may denote by $d+1$.

This is similar in spirit to the \textit{fully visible sigmoid belief network} (FVSBN) \cite{fvsbn}, which is in turn closely related to the \textit{neural autoregressive distribution estimator} (NADE) \cite{nade}. Both the FVSBN and NADE have been combined with RNNs and applied to algorithmic composition, in \cite{rh} and \cite{bl}, respectively. These methods employ a similar factorization to \eqref{eqn:tuplefactorisation}, but apply it to a binary indicator vector of dimension $d$, which has ones in those positions for which the musical note is sounding. Our approach is well suited to sparse problems (those in which $|\ovec| \ll d$), and also to unrolling in time in order to reduce the polyphonic music problem to a sequence of univariate prediction problems, as we explain in the next section. Indeed, to apply the unrolling in time idea of the next sub-section to an FVSBN type setup would require unrolling $d$ steps per chord, whereas as we shall see our approach need only unroll as many steps as there are notes in a given chord.
 
\subsection{Data Representation}
\label{sec:representation}

Following from the previous section, our idea is essentially to model each note in a piece of music sequentially given the previous notes, ordered first temporally and then, for notes struck at the same time, in ascending order of pitch. Similarly to the previous subsection, this defines a valid probability distribution which --- in contrast to the RNN-RBM but similarly to the RNN-NADE \cite{bl} --- is tractable by construction.\footnote{For example, there no known tractable means of computing the exact normalization constant of an RBM, as is required for computing say the test set likelihoods we report in \autoref{sec:experiments}.}

To achieve this, we employ the data representation of \autoref{fig:ty_maerz}. In this work, we assume the rhythmic structure, that is the note onset and offset times, is fixed. While we could model this as well, such a task is non-trivial and beyond the scope of this work. 

In \autoref{fig:ty_maerz}, the musical fragment is unrolled into an input matrix (first ten rows), a target row, and a lower bound row. At each time step, the LSTM is presented with a column of the input, and should predict the value in the target row, which is the midi number of the next note to be ``turned on''. The lower bound is due to the ordering by pitch --- notes with simultaneous onsets are ordered such that we can bound the value we are predicting, so we must incorporate this into the training loss function. 

Roughly speaking, the ``midi'' rows of the input indicate which notes are on at a given time. We can only allow one note to turn on at a time in the input, since we predict one note at a time, hence the second column has only pitch 70 turned on, and the algorithm should ideally predict that the next pitch is 74 (given a lower bound of 70), even though musically these notes occur simultaneously. Conversely, we can and do allow multiple pitches to turn off simultaneously (as in columns 5 and 10), since we are not predicting these --- the notes to turn off are dictated by the rhythmic structure (algorithmically, this requires some simple book keeping to keep track of which note(s) to turn off). 

The non-midi input rows 6 to 10 represent the rhythmic structure in a way which we intend to aid in the prediction. $\Delta t_{\textit{event}}$ is the duration of the note being predicted, $\Delta t_{\textit{step}}$ is the time since the previous input column, $\epsilon_\text{onset}$ is 1 if and only if we are predicting at the same time as in the previous column, $\epsilon_\text{offset}$ is 1 if and only if we are turning notes off at the same time as in the previous column, and $t$ represents the time in the piece corresponding to the current column (in practice we scale this value to range from 0 to 1). In the figure, the time columns are all in units of quarter notes.

This representation allows arbitrary timing information, and is not restricted to a uniform discretization of time as in many other works, \textit{e.g.} \cite{bl,moray}.  A major problem with the uniform discretization approach is that in order to represent even moderately complex music, the grid would need to be prohibitively fine grained, making learning difficult. Moreover, unlike the ``piano roll'' approaches we explicitly represent onsets and offsets, and are able to discriminate between, say, two eighth notes of the same pitch following one another as opposed to a single quarter note.

\subsection{Data Augmentation}
\label{sec:augmentation}

Data augmentation provides a simple way of encoding domain specific prior knowledge in any machine learning algorithm. For example, the performance of a generic kernel based classifier was advanced to the then state of the art in digit recognition by augmenting the training dataset with translations (in the image plane, for example by shifting all images by one pixel to the right), yielding the so-called virtual support vector machine algorithm \cite{virtualsvm}. There, the idea is that if the transformation leaves the class label unchanged, then the predictive power can be improved by having additional (albeit  dependent) data for training. It can also be rather effective to encode such prior knowledge directly in the model, however this is typically more challenging in practice \cite{lenet1}, and beyond the scope of the present work. 

Here we transpose our training data through all 12 (major or minor) keys, by transposing each piece by $n$ semi-tones, for $n=-6,-5,\ldots,4,5$. To the best of our knowledge this method has not been used in this context, with many authors instead transposing all pieces to some target key(s), such as C/A minor as in \cite{moray} or C/C minor as in \cite{bl}. Although simple and widely applicable to all approaches, we will see that augmenting in this way will lead to a uniformly substantial improvement in predictive power on all the datasets we consider. Moreover, by augmenting we can avoid the non-trivial sub-problem of key identification, which is arguably not even relevant for pieces that modulate. 

\section{Experiments}
\label{sec:experiments}

In this section, we first provide an overview of our experimental setup, before proceeding to demonstrate the efficacy of our approach. We proceed with the latter in two steps. First, in  \autoref{sec:bl} we verify the effectiveness of our model and data representation by comparing with previous approaches to the modelling of piano roll data. 
Then, in \autoref{sec:mid} we demonstrate our model on the new task of modelling non piano-roll data.

\subsection{Implementation Details and Experimental Setup}
\label{sec:implementation}

After representing the data in the form depicted in \autoref{fig:ty_maerz}, our algorithmic setup is similar to \cite{rnnregularization}, with two minor differences. The first is that some of the data columns do not have prediction targets (those with a dash in the target row of \autoref{fig:ty_maerz}). Here, we apply the usual LSTM recurrence but without any loss function contribution (in training) or output sampling (when generating samples from the trained model). The second difference is that the target value is lower bounded by a known value in some cases as described in \autoref{sec:representation}. The correct way to implement this is trivial: only those output layer logits whose indices satisfy the bound are considered at the output layer.
In line with the parameter naming conventions in \cite{rnnregularization}, we used two LSTM layers, an 800 dimensional linear embedding, a dropout ``keep'' fraction of 0.45, mini batches of size 50, and we limited the unrolling of the recurrence relation to 150 steps for computational reasons. We used the \textit{Adam} optimizer with the recommended default parameters \cite{adamoptimiser}. We optimized for up to 60 epochs, but terminated early if the validation likelihood did not improve (\textit{vs.} the best so far) while the training likelihood did improve (\textit{vs.} the best so far) four times in a row, at which point we kept the model with best validation score. This typically took 10 to 20 epochs in total, requiring up to around two days training for our largest problem. After training however, it only took around a second to sample an entire new piece of music, even with highly unoptimized code. Our implementation used Google's \textit{tensorflow} deep learning library,\footnote{\urllink{https://www.tensorflow.org}{www.tensorflow.org}} and we did most of the work on an NVIDIA Tesla K40 GPU.

\subsection{Benchmark Tests}
\label{sec:bl}

\begin{table*}
\caption{
\label{table:bl}
Mean test set log-likelihoods per time step, for the piano roll problems introduced in \cite{bl}.
}
\begin{center}
\begin{tabular}{lcccccc}
\\\thickhline
Model & Reference & Piano. & Nott. & Muse.  & JSB.  \\\thickhline
DBN-LSTM$^\text{\ref{foot:dbnlstm}}$ & \cite{dbnlstm} & -4.63 & -1.32 & -3.91 & -3.47 & \\
HMM (with chord annotations) & \cite{moray} & --- & --- & --- & -9.24 \\
TSBN & \cite{rh} & -7.98 & -3.67 & -6.81 & -7.48 \\
RNN-NADE & \cite{bl} & -7.05 & -2.31 & -5.60 & -5.56 \\
LSTM  & ours & -6.67 & -2.06 & -5.16 & -5.01 \\
LSTM (augmented) & ours &  -5.43 & -1.66 & -4.46 & -4.34 \\
LSTM (augmented + pooled) & ours & -4.94 & -1.57 & -4.41 & -4.45  \\
\hline
\end{tabular}
\end{center}
\end{table*}

\begin{table}
\caption{
\label{table:blcross}
Average test set log-likelihoods per time step for the piano roll problems introduced in \cite{bl}.
}
\begin{center}
\begin{tabular}{crcccc}
\multicolumn{2}{c}{\multirow{2}{*}{}} & \multicolumn{4}{c}{\textbf{Test}}\\
\cline{3-6} \\[-1.9ex]
\multicolumn{2}{c}{}                     & Piano. & Nott. & Muse. & JSB \\
\cline{3-6} \\[-1.9ex]
\multirow{5}{*}{\rotatebox[origin=c]{90}{\textbf{~~Train}}} & Piano. & -5.43 & -3.26 & -5.90 & -7.76 \\
                                      & Nott. & -15.28 & -1.66 & -17.12 & -18.40 \\
                                      & Muse. & -5.47 & -3.00 & -4.46 & -6.49 \\
                                      & JSB & -13.83 & -10.03 & -16.16 & -4.34 \\
\end{tabular}
%
%
\end{center}
\end{table}

Our first test setup is identical to \cite{bl}. In line with that work we report log likelihoods rather than the \textit{perplexity} measure which is more typical in NLP. Four datasets are considered, all of which are derived from midi files:\\
\textbf{Piano-midi.de} is a collection of relatively complex classical pieces transcribed for piano \cite{poliner}.\\
\textbf{Nottingham} is a collection of folk tunes with simple melodies and block chords.\footnote{
\urllink{http://tinyurl.com/z2uofrp}{ifdo.ca/\textasciitilde{}seymour/nottingham/nottingham.html}
} \\
\textbf{MuseData} is a collection of high quality orchestral and piano pieces from CCARH.\footnote{\urllink{http://www.musedata.org}{www.musedata.org}}\\
\textbf{J. S. Bach} is a corpus of chorales harmonized by J.S. Bach, with splits taken from \cite{moray}.

To compare with previous results \cite{bl,rh} we used the pre-processed data provided in piano roll format, which is a sequence of sets of notes lying on a uniform temporal grid with a spacing of an eighth note, and pre-defined train/test/validation splits. To adapt our more general model to this problem, we added an ``end'' symbol along with each set of notes, as per \autoref{sec:reduction}. We also omitted the $t$ row of \autoref{fig:ty_maerz}, which contains a form of ``forward information''. In this way, our model solved precisely the same modelling problem as \cite{bl}, and our likelihoods are directly comparable.

The results in \autoref{table:bl} include the best model from \cite{bl}, as well as that of \cite{rh}. The \textit{LSTM} row corresponds to our basic model, which performs encouragingly, outperforming the previous methods on all four datasets. The \textit{LSTM (augmented)} row uses the data augmentation procedure of \autoref{sec:augmentation}, which proves effective by outperforming the basic model (and all previous approaches) convincingly, on all datasets. The only exception is the DBN-LSTM model\footnote{\label{foot:dbnlstm} 
However, while not stated in \cite{dbnlstm}, the DBN-LSTM likelihoods are well known merely to be optimistic bounds on the true values.} Finally, in addition to the augmentation, we pooled the data from (the training and validation splits of) all four datasets, and tested on each dataset, in order to get an idea of how well the model can generalize across the different data sets. The difference here is minor in all cases but piano-midi.de, where we see further improvement, which is interesting given that piano-midi.de appears to be the most complex dataset overall. There is a small deterioration of this \textit{LSTM (augmented + pooled)} model performance on the J.S. Bach test set, however \autoref{table:blcross}, which is a cross table of performance for training on one dataset and testing on another, shows that this is not the complete picture. Indeed,  models trained and tested on different datasets unsurprisingly do far worse than our pooled data model. This suggests that our model with pooling and augmentation as the best overall for algorithmic composition.

\subsection{A New Benchmark Problem}
\label{sec:mid}

\begin{table*} 
\caption{
\label{table:blmid}
Mean test set log-likelihoods per note for the midi data of \cite{bl}.
}
\begin{center}
\begin{tabular}{lccccc}
\\\thickhline
Model & Piano. & Nott. & Muse.  & JSB.  \\\thickhline
LSTM  & -2.99 & -1.18 & -1.71 & -1.32 \\
LSTM (augmented) &  -2.28 & -0.99 & -1.33 & -1.07 \\
LSTM (augmented + pooled) & -2.14 & -0.99 & -1.34 & -1.25 \\
\hline
\end{tabular}
\end{center}
\end{table*}

In the previous section, we adapted our model to handle the case where time is uniformly discretised, but the number of notes sounding at each time step is unknown.  In this section, we use the original midi file data to construct a different, arguably more musically relevant problem. The data, which we have preprocessed as described in the supplementary material, and made available online$^\text{\ref{FOOT:supplementary}}$ is now in the form of sets of tuples indicating the onset time, offset time, midi note number, and part number (\textit{i.e.} midi track number) of each event in the original midi file. We model the midi note names conditional on the rhythmic structure (onset and offset times). We also ignore the part numbers, although future work should utilize these, given that individual parts in a piece of music should ideally have a coherence of their own. 

Before modelling the above data, we applied a simple preprocessing step in order to clean the timing information. This step was necessary due to the imprecise timing information in some of the source midi files, and would have no effect on ``clean'' midi data derived from musical scores. In particular, we quantized the onset and offset times to the nearest 96th of a quarter note, unless this would collapse a note to zero duration in which case we left the onset and offset times unchanged. This was especially helpful for the case of the \textit{piano-midi.de} data, where the onset and offset times often occur extremely close to one another. There are two distinct reasons why we apply this pre-processing step:
\begin{enumerate}
\item
By quantizing the offset times, we reduce the number of columns in our data representation, since more notes can ``turn off'' simultaneously in a single input column. See \autoref{fig:ty_maerz} and the corresponding description of the data representation in \autoref{sec:representation}. This in turn effectively allows the LSTM to model longer range temporal dependencies.
\item 
By quantizing the onset times, our ordering technique (see subsections \ref{sec:reduction} and \ref{sec:representation}), which orders first by time and then by pitch, will more consistently represent those onsets which are close enough to be considered practically simultaneous.
\end{enumerate}

The results in \autoref{table:blmid} are now in the form of mean log-likelihood per note, rather than per time step, as before. In the setup of \cite{bl}, a uniformly random prediction would give a log likelihood of $-\log(2^d)$ (we assume no unison intervals) where $d$ is the number of possible notes. For $d=88$ notes (spanning the full piano keyboard range of A0 to C8) as in \cite{bl}, this number is approximately $-61$. For our new problem however, a uniformly random prediction will have a mean log-likelihood per note of $-\log(d)$ which is approximately $-4.47$ for $d=88$. Overall then, the results in \autoref{table:blmid} are qualitatively similar to the previous results, after accounting for the expected difference of scale. 

Note that it is not practical to compare with the previous piano-roll algorithms in this case. It turns out that the greatest common divisor of note durations in our data dictates a minimum temporal resolution of 480 ``ticks'' per quarter note, rendering those algorithms impractical (note that the piano roll format used in the previous work and in \autoref{sec:bl} used a resolution of an eighth note, or two ticks per quarter note).

\begin{figure}
\centering
  \includegraphics[width=0.75\textwidth]{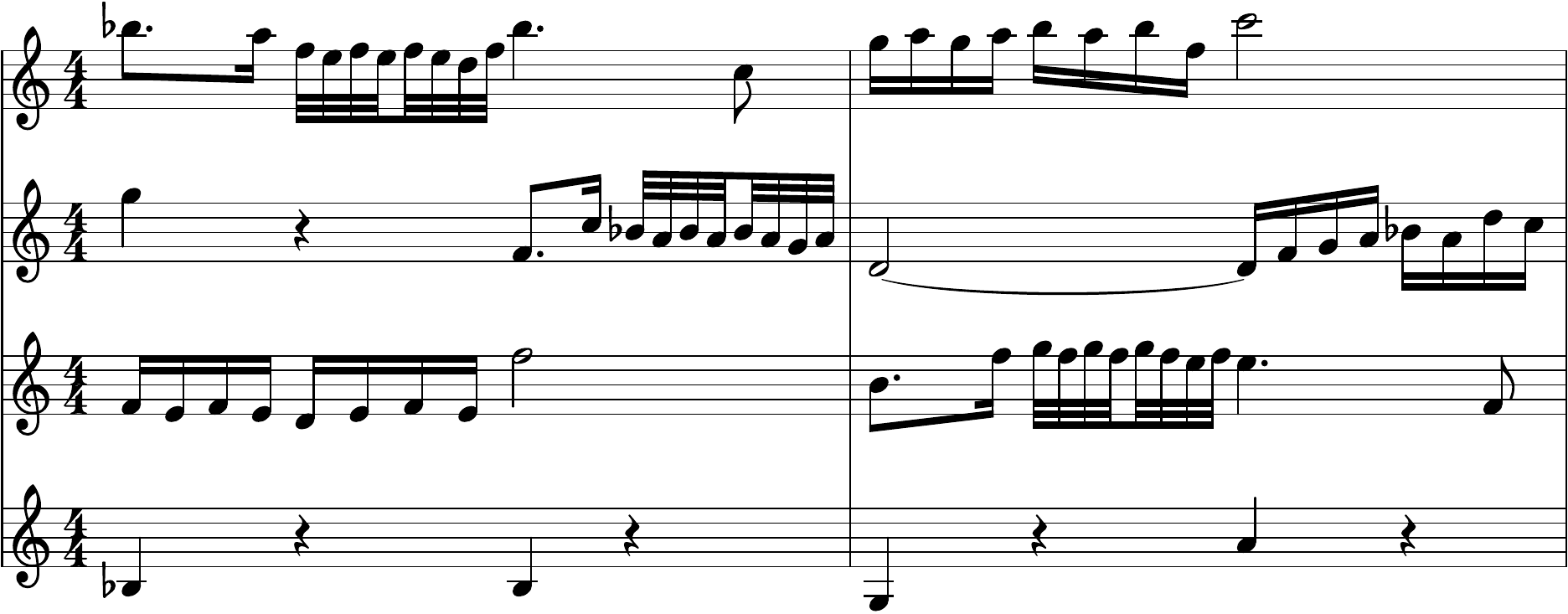}
\caption{ 
\label{fig:sample}
Two bars of the model's output, featuring a relatively complex rhythmic sequence which occurs 52 seconds into the audio sample included in the supplementary material.
}
\end{figure}

Importantly however, we can generate music with an arbitrarily complex rhythmic structure. We generated a large number of example outputs by sampling from our model given the rhythmic structure from each of the test set pieces contained in the MuseData corpus.\footnote{\label{FOOT:supplementary} Our data and output audio samples:  \urllink{\walderurl}{\walderurl}} 
We also included a single sample output in the supplementary material, two bars of which are notated in \autoref{fig:sample}.

\begin{figure}
  \centering
  \includegraphics[width=0.5\textwidth]{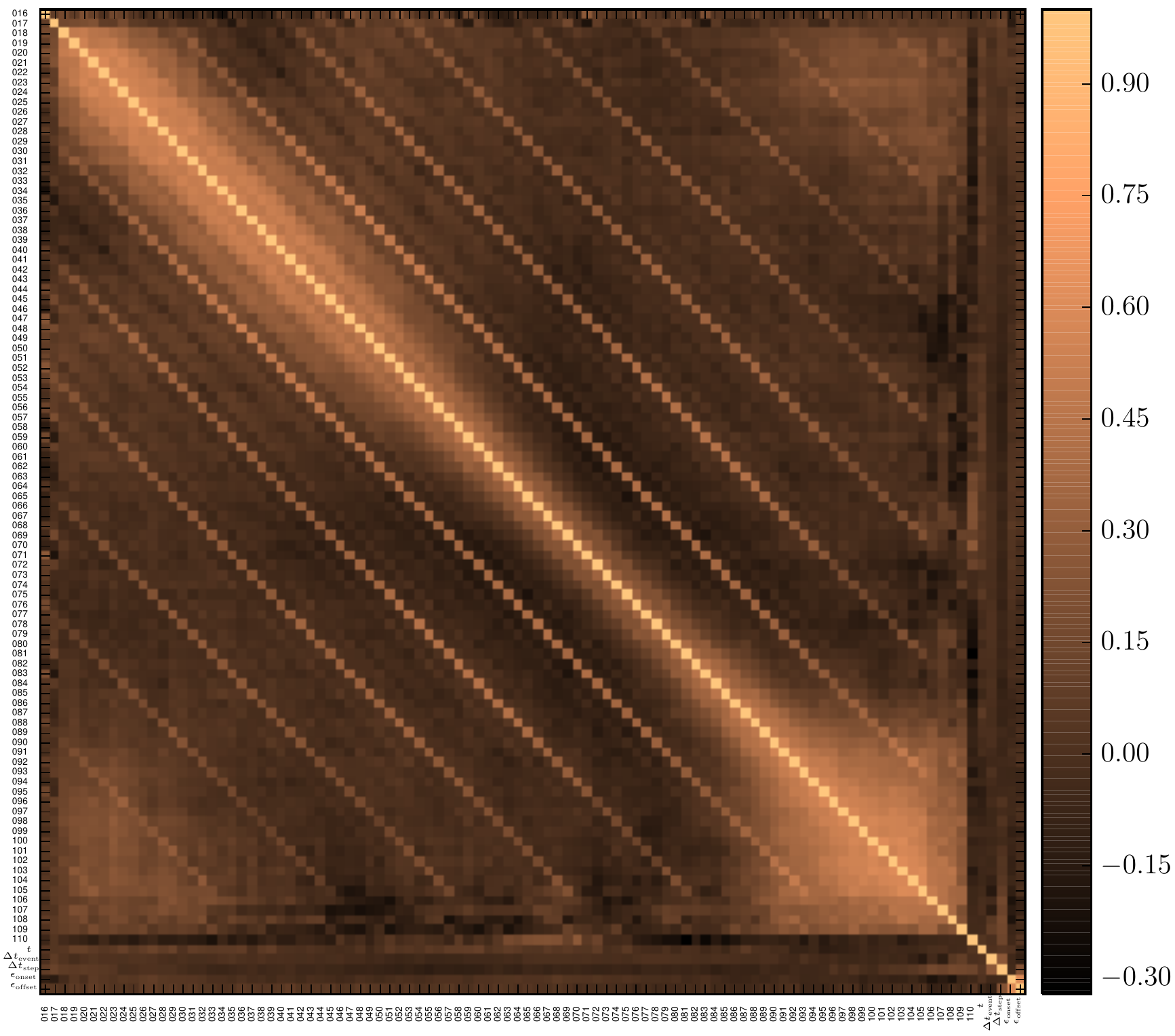}
\caption{
\label{fig:embedding}
Correlations between the columns of the learned \textit{embedding matrix}, which maps inputs to higher dimensional ``semantic'' representations. The row/column labels can be seen by zooming in on an electronic copy: the last 5 rows/columns correspond to the last five rows of the input (upper) part of Figure \ref{fig:ty_maerz} (b), the remaining rows/columns correspond to sequential midi numbers. The off-diagonal stripes suggest that the model has discovered the importance of the octave. 
}
\end{figure}

We also attempted to visually identify some of the structure we have captured. For the pooled and augmented model, we visualized the embedding matrix which linearly maps the raw inputs into what in NLP would be termed a latent semantic representation. A clear structure is visible in \autoref{fig:embedding}. The model has discovered the functional similarity of notes separated by an interval of one or more octaves. This is interesting given that all notes are orthogonal in the raw input representation, and moreover that the training was  unsupervised. In NLP, similar phenomena have been observed \cite{glove}, as well as more sophisticated relations such as (in embedding space) ``\textit{Queen} - \textit{King} + \textit{brother} $\approx$ \textit{sister}''. We tried to identify similar higher order relationships in our embedding, by comparing intervals, analyzing mapped counterparts to the triads of the circle of fifths, \textit{etc.}, but the only clear relationship we discovered that of the octave, described above. This does not mean the model hasn't identified these relationships however, since the embedding is only the first layer in a multi-layer model.

\section{Summary and Outlook}
\label{sec:summary}


We have presented a model which reduces the problem of polyphonic music modelling to one of modelling a temporal sequence of univariate categorical variables. By doing so, we were able to apply state of the art neural language models from natural language processing. We further obtain a relatively large improvement in performance by augmenting the training data with transpositions of the pieces through all musical keys --- a simple but effective approach which is directly applicable to all algorithms in this domain. Finally, we obtain a moderate additional improvement by pooling data from various sources. An important feature of our approach is its generality in terms of rhythmic structure --- we can handle any fixed rhythmic structure and are not restricted to a uniform discretization of time as in a piano roll format. This lead us to derive a new benchmark data set, which we believe will be useful for further advancing the state of the art in algorithmic composition. Future work could 1) model the rhythmic information, rather than conditioning on it as we do here, 2) provide a mechanism for user constraints on the output, as has been done with Markov models by applying branch and bound search techniques \cite{pachetmarkovconstraints2011}, or 3) introduce longer range dependencies and motif discovery using attentional models (see \textit{e.g.} \cite{lstmattention}).

\section{Acknowledgments}

We gratefully acknowledge NVIDIA Corporation for donating the Tesla K40 GPU used in this work.

\bibliography{walder}
\bibliographystyle{unsrt}
\end{document}


\maketitle

\appendix

\section{Data}

\subsection{Introduction}

In this appendix we provide an overview of the symbolic music datasets we offer in pre-processed form\footnote{The data is available for download here: \urllink{\walderurl}{\walderurl} \label{foot:dataurl}}. Note that the source of these datasets is the same set of midi files used in \cite{bl}, which also provides pre-processed data. That work provided ``piano roll'' representations, which essentially consist of a regular temporal grid (of period one eighth note) of on/off indicators for each midi note number. While the piano roll is an excellent simplified music format for early investigations into symbolic music modelling, it does have several limitations, as discussed in the main text. To name one such limitation, the piano roll format does not explicitly represent note endings, and therefore cannot differential between, say, two successive eighth notes, and a single quarter note.

To address these limitations, we have extracted additional information from the same set of midi files. Our goal is to represent the performance (or sounding) of notes by when they begin and end, rather than whether they are sounding or not at each time on a regular grid. The representation we adopt consists of sets of five-tuples, each of which describes the sounding (and termination) of a single note. Each such note event is represented by the following integers:
\begin{itemize}
\item piece number (corresponding to a midi file),
\item track (or part) number, defined by the midi channel in which the note event occurs,
\item midi note number, ranging 0-127 according to the midi standard, and 22-105 inclusive for the data we consider here,
\item note start time, in ``ticks'', (2400 ticks = 1 beat = one quarter note),
\item note end time, also in ticks.
\end{itemize}
We split the pieces into the same three (train/validation/test) splits as used in \cite{bl}.

Please refer to the data archive itself$^\text{\ref{foot:dataurl}}$ for a detailed description of the format. 

A summary of the four datasets is provided in \autoref{table:datasummary}.

\subsection{Preprocessing}

We applied the following processing steps and filters to the raw midi data.
\begin{itemize}
\item Combination of piano ``sustain pedal'' signals with key press information to obtain equivalent individual note on/off events. 
\item Removal of duplicate/overlapping notes which occur on the same midi channel (while not technically allowed, this still occurs in real midi data due to the permissive nature of the midi file format). Unfortunately, this step is ill posed, and different midi software packages handle this differently. Our approach involves processing notes sequentially in order of start time, and ignoring those note events that overlap a previously added note event.
\item Removal of midi channels with less than two note events (these occurred in the MUS dataset, and were always information tracks containing authorship information and acknowledgements, \etc).
\item Removal of percussion tracks. These occurred in some of the Haydn symphonies and Bach Cantatas contained in the MUS dataset. It is important to filter them as the percussion instruments are not pitched, and hence the midi numbers in these tracks are not comparable with those of pitched instruments, which we aim to model.
\item Re-sampling of the timing information to a resolution of 2400 ticks per quarter note, as this is the lowest common multiple of the original midi file resolutions (see \autoref{table:datasummary}).
\end{itemize}

\subsection{Exploratory analysis}

We provide some basic exploratory plots in figures \ref{fig:PMD}--\ref{fig:MUS}. 

The \textbf{Note Distribution} and \textbf{Number of Notes Per Piece} plots are self explanatory.

Note that the \textbf{Number of Parts Per Piece} (lower left sub figure) is fixed at one for the entire JSB dataset. This is due to an unfortunate lack of midi track information in these files, many of which are in fact four part harmonies. The pieces in the NOT dataset feature either one part (in the case of pure melodies) or two (in the case of melodies with associated chord accompaniments). The PMD dataset features up to six parts (for a three-part Bach fugue in which left and right hands are tracked separately). MUS features up to 27 parts (for Bach's \textit{St. Matthew's Passion}). Note that this information is for future reference only --- we have yet to take advantage of the part information in our models.

The least obvious sub-figures are those on the lower-right labeled \textbf{Peak Polyphonicity Per Piece}. Polyphonicity simply refers to the number of simultaneously sounding notes, and this number can be rather high. For the PMD data, this is mainly attributable to musical ``runs'' which are performed with the piano sustain pedal depressed, for example in some of the Liszt pieces. For the MUS data, this is mainly due to the inclusion of large orchestral works which feature many instruments.

\bibliography{walder}
\bibliographystyle{unsrt}

\begin{table*}
\begin{center}
\begin{tabular}{ccccc}
\\\thickhline
Dataset & Long Name & Source & Total Pieces & Midi Resolution
\\\thickhline
PMD & \texttt{piano-midi.de} & \cite{poliner,bl} & 124 & 480 \\
\hline
JSB & J.S Bach Chorales & \cite{moray,bl} & 382 & 100 \\
\hline
MUS & MuseData & \cite{musedata,bl} & 783 & 240 \\
\hline
NOT & Nottingham & \cite{nottinghamdata,bl} & 1037 & 480 \\
\hline 
\end{tabular}
\end{center}
\caption{
\label{table:datasummary}
A summary of the datasets used in this study.
}
\end{table*}
\begin{tabular}{|c|c|c|}
\end{tabular}

\newcommand{\dataplot}[2]{%
\begin{figure*}%
\begin{center}%
  \includegraphics[width=0.45\textwidth,page=1]{#1}%
  \includegraphics[width=0.45\textwidth,page=2]{#1}%
  \\
  \includegraphics[width=0.45\textwidth,page=3]{#1}%
  \includegraphics[width=0.45\textwidth,page=4]{#1}%
  \caption{%
  \label{fig:#2}%
  Summary of the #2 dataset.%
  }%
\end{center}%
\end{figure*}%
}

\clearpage 

\dataplot{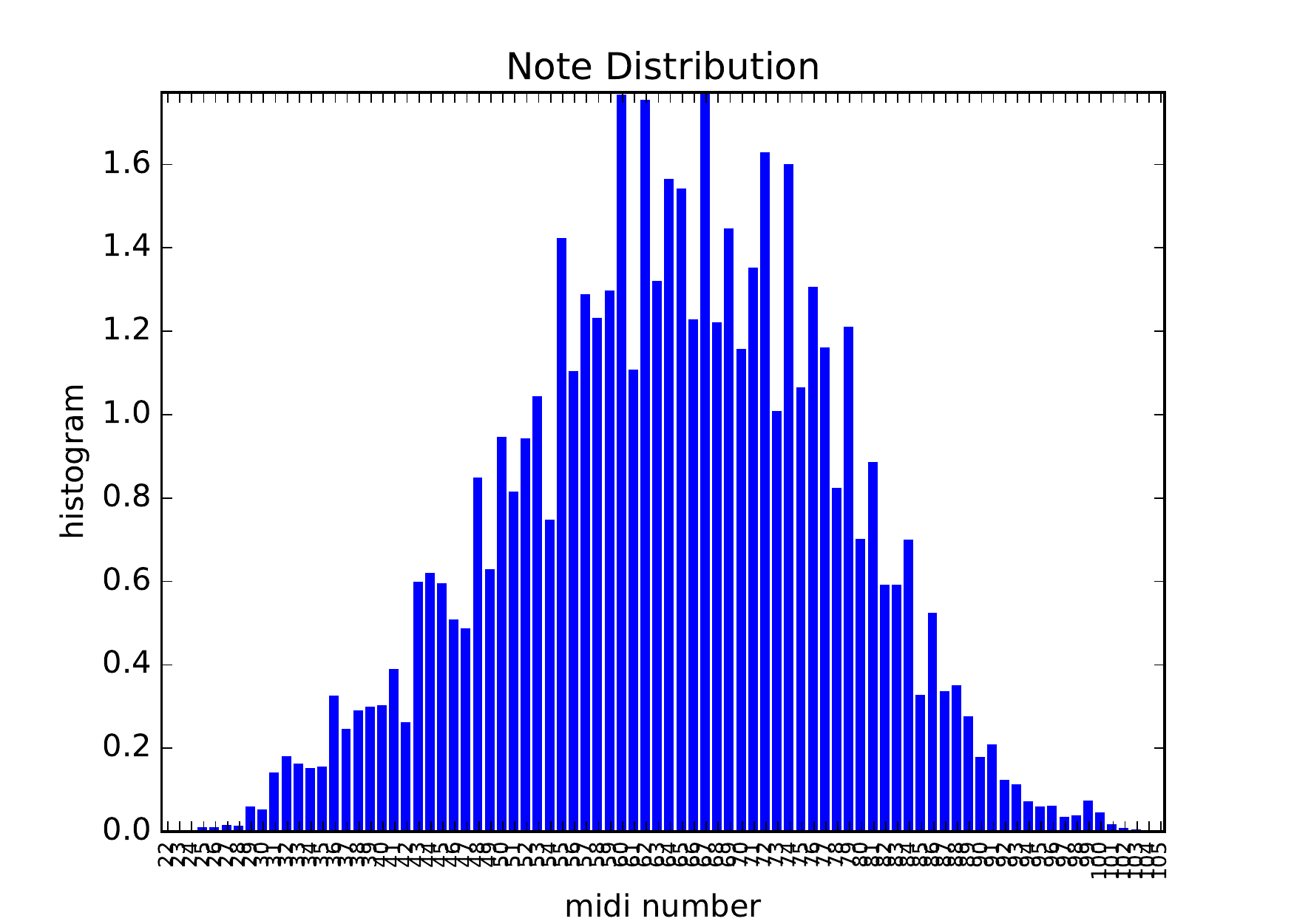}{PMD}
\dataplot{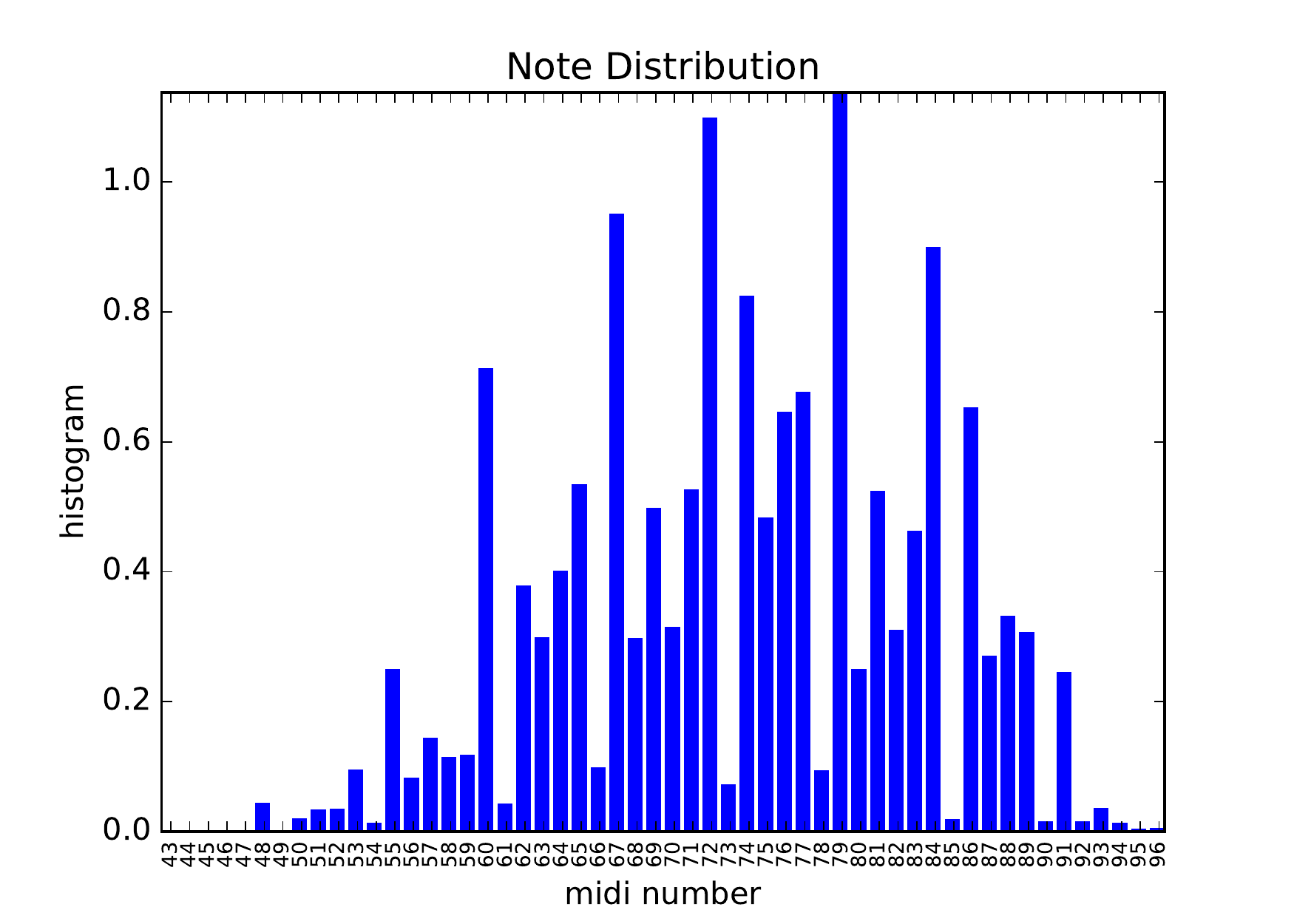}{JSB}
\dataplot{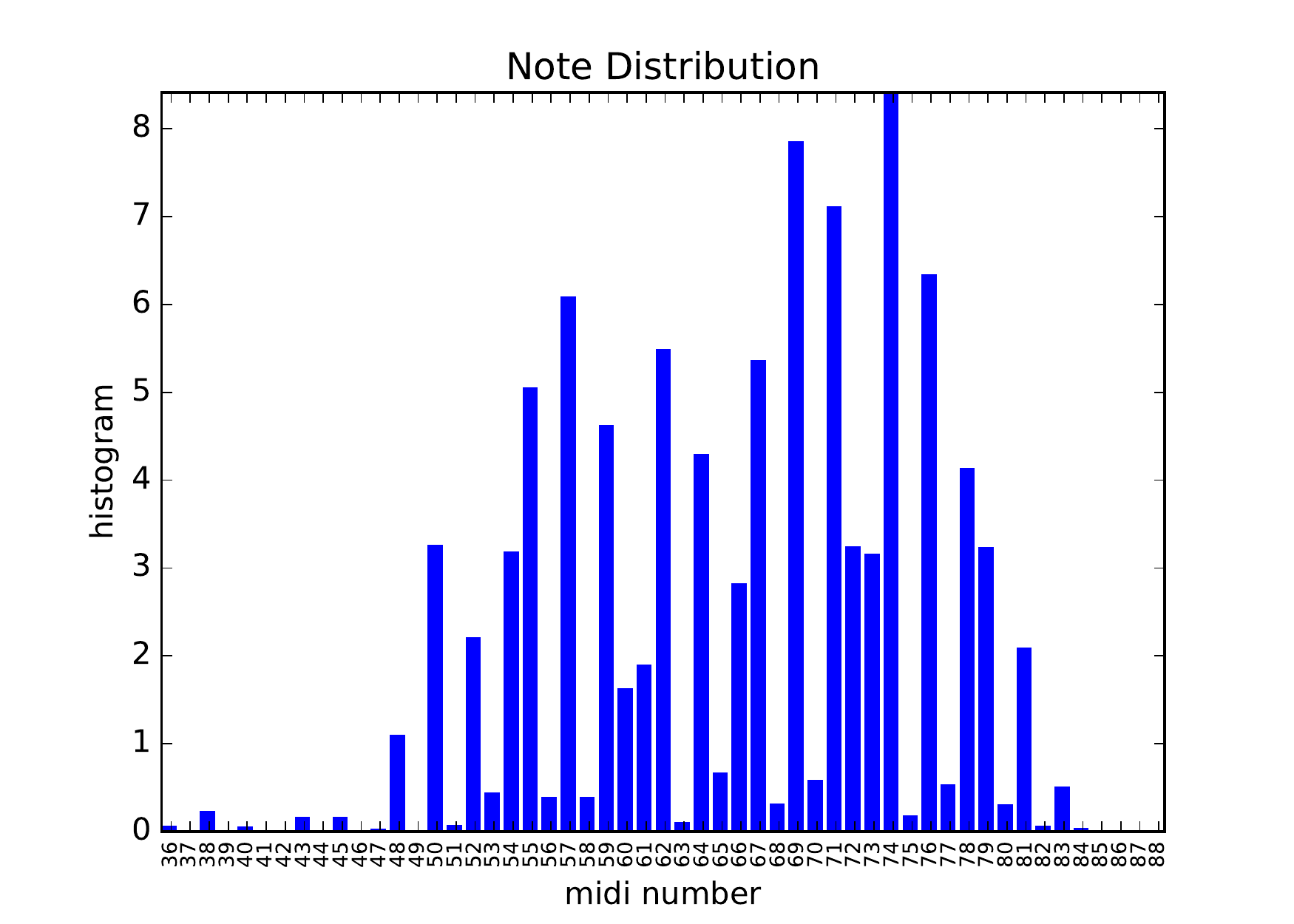}{NOT}
\dataplot{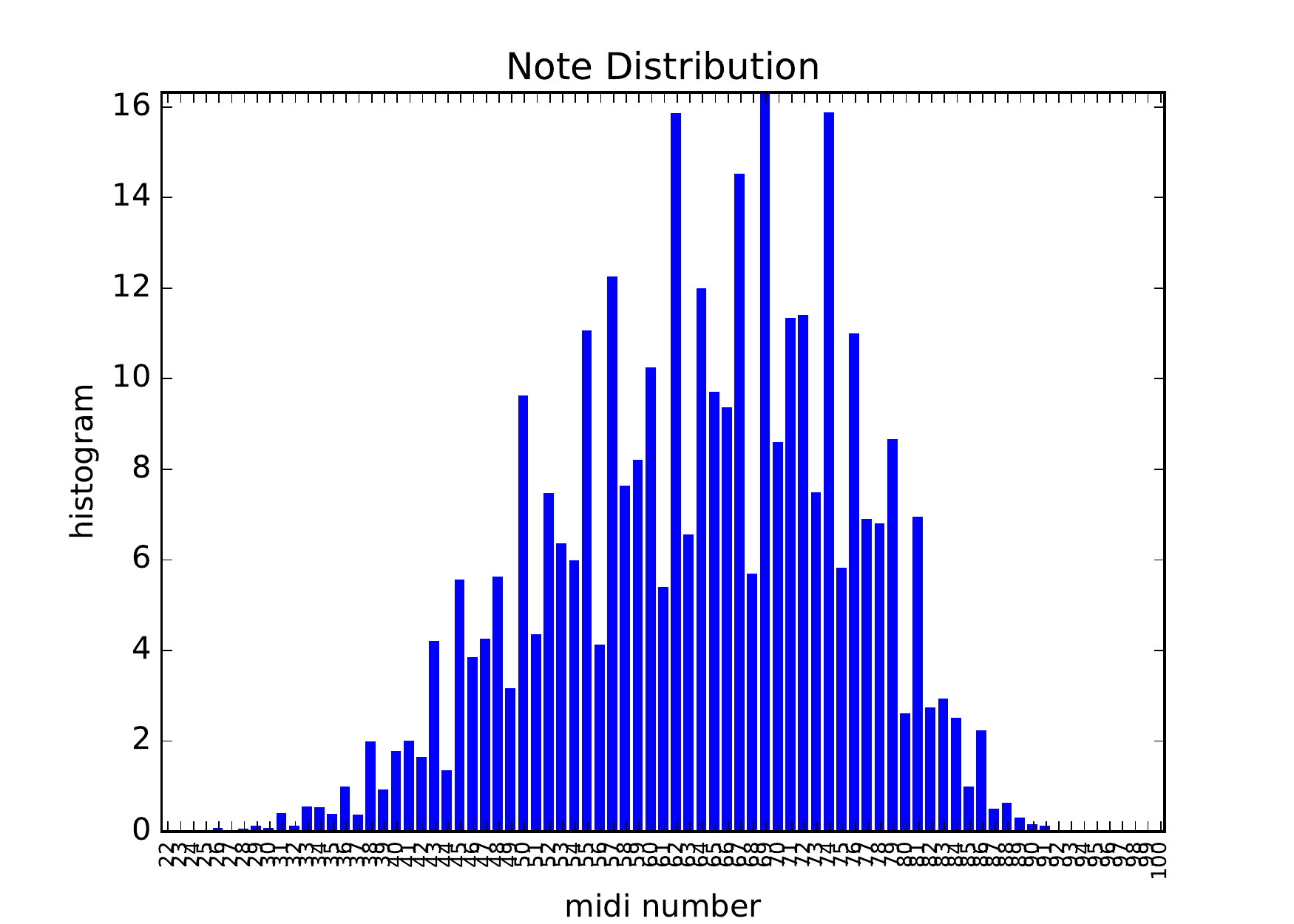}{MUS}